\documentclass{article}

\PassOptionsToPackage{numbers, compress}{natbib}
\usepackage[preprint]{neurips_2021}

\usepackage[utf8]{inputenc} 
\usepackage[T1]{fontenc}    
\usepackage{hyperref}       
\usepackage{url}            
\usepackage{booktabs}       
\usepackage{amsfonts}       
\usepackage{nicefrac}       
\usepackage{microtype}      
\usepackage{xcolor}         
\usepackage[draft,nomargin,inline,author=]{fixme}
\fxsetup{theme=color}
\definecolor{fxwarning}{rgb}{0.8,0.0000,0.0000}
\usepackage{comment}
\usepackage{amsmath}
\usepackage{graphicx}
\usepackage{algorithm}
\usepackage[noend]{algpseudocode}
\usepackage{siunitx}
\usepackage{caption}
\usepackage{subcaption}

\title{A Photonic-Circuits-Inspired Compact Network: Toward Real-Time Wireless Signal Classification at the Edge}

\author{%
  Hsuan-Tung Peng\thanks{Corresponding author} \\
  Princeton University\\
  \texttt{hpeng@princeton.edu} \\
  \And
  Joshua Lederman \\
  Princeton University \\
  \texttt{joshuacl@princeton.edu} \\
  \AND
  Lei Xu \\
  Princeton University \\
  \texttt{leixu@princeton.edu} \\
  \And
  Thomas Ferreira de Lima \\
  Princeton University \\
  \texttt{tlima@princeton.edu} \\
  \And
  Chaoran Huang\thanks{Currently with Department of Electronics Engineering, the Chinese University of Hong Kong} \\
  Princeton University \\
  \texttt{chaoranh@princeton.edu} \\
  \And
  Bhavin Shastri \\
  Queen's University $\&$ \\
  Vector Institute \\
  \texttt{bhavin.shastri@queensu.ca} \\
  \And
  David Rosenbluth \\
  Lockheed AI center \\
  \texttt{rosenbluthd@gmail.com} \\
  \And
  Paul Prucnal \\
  Princeton University \\
  \texttt{prucnal@princeton.edu}
}
\begin{document}

\maketitle

\begin{abstract}
  Machine learning (ML) methods are ubiquitous in wireless communication systems and have proven powerful for applications including radio-frequency (RF) fingerprinting, automatic modulation classification, and cognitive radio. However, the large size of ML models can make them  difficult to implement on edge devices for latency-sensitive downstream tasks. In wireless communication systems, ML data processing at a sub-millisecond scale will enable real-time network monitoring to improve security and prevent infiltration. In addition, compact and integratable hardware platforms which can implement ML models at the chip scale will find much broader application to wireless communication networks. Toward real-time wireless signal classification at the edge, we propose a novel compact deep network that consists of a photonic-hardware-inspired recurrent neural network model in combination with a simplified convolutional classifier, and we demonstrate its application to the identification of RF emitters by their random transmissions. With the proposed model, we achieve 96.32\% classification accuracy over a set of 30 identical ZigBee devices when using 50 times fewer training parameters than an existing state-of-the-art CNN classifier. Thanks to the large reduction in network size, we demonstrate real-time RF fingerprinting  with 219 \si{\micro\second} latency using a small-scale FPGA board, the PYNQ-Z1.
\end{abstract}
\section{Introduction}\label{sec:intro}

Wireless communication systems have been embracing machine learning (ML) and neural networks in order to improve security, spectral utilization and operational efficiency. When ML is adopted for wireless signal classification, the received radio-frequency (RF) signals are digitized, pre-processed, and further fed into a neural network for specific tasks, such as emitter identification ~\cite{Merchant2018DeepLF}, modulation format classification~\cite{ModClassification2016}, and position mapping~\cite{PositionMapping2008}. With optimization, the current deep neural networks have consistently increased the classification accuracy in various application scenarios. However, the existing ML solutions employing conventional deep neural networks are deficient in addressing two important requirements: (1) real-time classification with (2) small size and power for the mobile edge. Among the various ML applications in wireless systems, we use RF fingerprinting as a typical example to present our algorithms, implementation and results in tackling the challenge of real-time wireless signal classification with small size and power.

Current state-of-the-art RF fingerprinting techniques require pre-existing knowledge of RF engineering~\cite{Chen2017IdentificationOW} or large deep neural networks~\cite{Merchant2018DeepLF,Yu2019ARR,Jian2020DeepLF} and process radio signals offline. Although these approaches have achieved noteworthy results in demonstrating the effectiveness of ML for RF signal processing and classification, they face significant challenges during implementation in real-time at the network edge and are not scalable due to latency introduced by the digital electronic hardware and the large size of the network architectures.

In this work, we present a compact neural network model that consists of a photonics-inspired recurrent neural network (PRNN) to efficiently transform temporal data at the front end and a convolutional neural network (CNN) at the back end to perform classification. With the PRNN, we can learn more representative features and reduce the number of parameters of the CNN classifier. Thanks to the significant size reduction  of the neural network classifier, we are able to implement the proposed PRNN and CNN model on a low-cost and small-scale FPGA and achieve sub-millisecond RF ZigBee signal classification.


The main contributions of this paper are:
\begin{enumerate}
  \item We propose a PRNN model to learn more expressive features and enable a more compact CNN classifier for RF fingerprinting~\cite{Merchant2018DeepLF}.
  \item We implement the proposed deep neural network on a small-scale FPGA to demonstrate real-time processing for RF fingerprinting.
  \item We evaluate hardware performance using several metrics including energy, latency and throughput. 
\end{enumerate}
The rest of the paper is organized as follows: in the next section, we introduce our problem settings for RF fingerprinting and discuss a previous approach to tackle this problem.
In Section \ref{sec:PRNN-CNN}, we present our PRNN-CNN model. In Section \ref{sec:exp}, we show  classification results for 30 ZigBee devices and its performance on an FPGA. Conclusions are drawn in the last section.

\section{Preliminary: RF fingerprinting}
Internet of things (IoT) devices without consistent human possession can be vulnerable to authentication-related threats, such as spoofing attacks in which digital IDs are cloned. \emph{RF fingerprinting} is one method to prevent such attacks, involving the identification of specific devices by their unique emission characteristics originating from manufacturing imperfections. The variation in behavior among devices is statistical in nature and imparts a device-specific signature to the transmitted signals, allowing for RF identification. However, the device signature is usually hidden by the much stronger signal carrying the information content of the transmission. Extracting the sparsely distributed and low signal-to-noise ratio (SNR) features and achieving high classification accuracy are the main challenges of RF fingerprinting.
\subsection{Prior work}\label{sec:prior}
Recently, a number of machine learning based methods have been proposed for RF fingerprinting. ~\citet{Youssef2018MachineLA} investigated the use of machine learning strategies including conventional deep neural nets (DNNs), CNNs, support vector machines (SVMs), and DNNs with multi-stage training to the classification and identification of RF Orthogonal Frequency-Division Multiplexing (OFDM) packets in the time domain. ~\citet{Das2018ADL} presented a long short-term memory (LSTM) network that exploits temporal correlation between the I/Q streams of wireless signals to identify low-power transmitters and evaluate system performance in various adversarial settings. ~\citet{Merchant2018DeepLF} demonstrated a  multi-layer CNN to classify error signals in the transmitted waveforms, where the trivial features such as carrier frequency and phase offsets are removed.
Although these methods show reasonably good performance, due to the size of the networks low-latency RF fingerprinting cannot be achieved.

\subsection{Problem formulation}

We define our RF fingerprinting problem as the classification of demodulated and digitized I/Q signals from 30 identical Digi XBP24CZ7SITB003 ZigBee Pro devices with a carrier frequency of 2.405 GHz. Each transmission has a 32-byte random payload and a 2-byte checksum, and the data is provided by the Naval Research Laboratory (NRL)~\cite{Merchant2018DeepLF}.  The transmissions contain features such as carrier frequency and phase offsets that can be used for trivial classification but are susceptible to an adversary's spoofing of the local oscillator of a transmitter. Therefore, to prevent this type of attack and further enhance the security of RF fingerprinting, we pre-process the raw demodulated I/Q samples from the transmissions to generate {\it{residual data}}, which is defined as follows:

\begin{itemize}
\item {\it{Residual Data}} = Received Raw Data $-$ $f_{\omega,\phi}$(Ground Truth Data)
\end{itemize}
Here, $\omega$ and $\phi$ are the carrier frequency and phase offset of the transmission, respectively, $f_{\omega, \phi}$ is a function that applies the frequency and phase offset $\omega,\phi$ to the ground truth data, and the ground truth data is the signal that we regenerate after decoding the transmission. To derive $\omega, \phi$ and residual data, we follow the ZigBee data recovery steps proposed by \citet{Merchant2018DeepLF}. The details of residual processing are provided in Appendix A. After residual data processing, the trivial signatures (carrier frequency and phase offset) are removed from the received raw data. Furthermore, by subtracting the ground truth signal, the model can detect the device-specific features that are usually hidden within the received raw signal.

The final dataset is the residual data consisting of 34 bytes per transmission and about 1,120 transmissions for each device. We split $80\%$ of the transmissions for the training dataset, $10\%$ for the development dataset, and $10\%$ for the test dataset.

\section{PRNN-CNN model for RF  fingerprinting}\label{sec:PRNN-CNN}
In this work, we propose a compact neural network model to classify on the steady state of the transmission and perform a benchmark comparison to the model proposed by Ref.~\cite{Merchant2018DeepLF}, with which we share the same source data. 
For the rest of this manuscript, we use \emph{"NRL CNN"} to refer to the baseline model proposed by the Naval Research Laboratory~\cite{Merchant2018DeepLF}, and call our proposed model the \emph{"PRNN-CNN"}.

\subsection{I/O settings}\label{sec:data_form}
The original residual data has 2 channels (I/Q) and a total of 34 bytes (17,408 samples). In the same manner as Ref.~\cite{Merchant2018DeepLF}, we define a two-byte segment of the transmission as the data unit for the classifier. As such, each transmission has 17 data units, where each unit is a two-channel time series with 1,024 steps. We denote the $n$-th data unit of a transmission as $Z_n=[Z_n^0,...,Z_n^i,..., Z_n^{1023}],$ $\forall n \in 0,1,...,16$. Here, at each time step, $Z_n^i=[I_n^i,Q_n^i]^T$ is the $i$-th sample in the data unit. We treat the consecutive 32 samples of each channel as features and reshape the input from (2,1024) to (64,32). The reshaped data unit can be represented as
\begin{align}
    X_n & =[X_n^0, ..., X_n^i, ..., X_n^{31}], \nonumber \\
    X_n^i & =[I_n^{32i}, I_n^{32i+1}, ... I_n^{32i+31}, Q_n^{32i}, Q_n^{32i+1}, Q_n^{32i+31}]^T,  \forall i \in 0,1,...,31
\end{align}

The model will take each segment of data $Y_n$ at once and output a log probability $\log\vec{P}_n =[\log P_n^1,...,\log P_n^k,..., \log P_n^{30}], \forall k \in 1,2,...,30$, where $\log P_n^k=\log P(C_k|X_n)$, and $C_k$ is the $k$-th transmitter.
The overall classification result is determined by the accumulated log probability of $N$ data units of each transmission, i.e. $\log \vec{P}=\sum_{n=1}^N \log \vec{P}_n$. Here, $N$ can be chosen from 1,2...,17.
\subsection{PRNN model}

The proposed PRNN model is inspired by silicon photonic neural networks~\cite{Tait:2017aa,Tait:2019}.
Recently silicon photonics-based integrated circuits have found important applications in machine learning due to their high bandwidth, low latency, low power consumption, and parallel processing capability ~\cite{Lima2019MachineLW}. Photonic RNNs, implemented using a broadcast-and-weight system configured by microring weight banks, have been shown to be isomorphic to continuous-time RNN (CTRNN) models with application to ultrafast information processing for control and scientific computing~\cite{Tait:2017aa}.

Here, we propose using the photonic hardware compatible RNN model~\cite{Lima2019MachineLW,Tait:2017aa} to extract the temporal features in a transmission and emulating this model on an FPGA to achieve real-time processing. In the future, this RNN model can be further realized on a fully-integrated photonic neural network to improve the latency of the RF fingerprinting system.

The dynamics of a photonic recurrent neural network can be described by the following equation~\cite{Tait:2017aa}:
\begin{align}
    & \frac{d\vec{s}}{dt}  =\frac{-\vec{s}}{\tau}+{\bf{W}}\vec{y}(t) \\
    & \vec{y}(t) = \sigma(\vec{s}(t))
\end{align}
where $\vec{s}$ is the neuron's state, $\vec{y}$ is output signal, $\tau$ is the time constant of the photonic circuit, ${\bf{W}}$ is the photonic weight, and $\sigma(\cdot)$ is the transfer function of the silicon photonic modulator neurons~\cite{Tait:2019,Huang2020DemonstrationOP}. It is worth noting that the nonlinear transfer function can be expressed as Lorentzian function, $\sigma(x)=x^2/(x^2+(ax+b)^2)$, where $a, b$ are constants. In this section, we treat this dynamical system as an RNN model and leave the details of physics and its relation with silicon photonic circuits to Appendix B.

\begin{figure}[ht!]
    \centering
    \includegraphics[width=\linewidth]{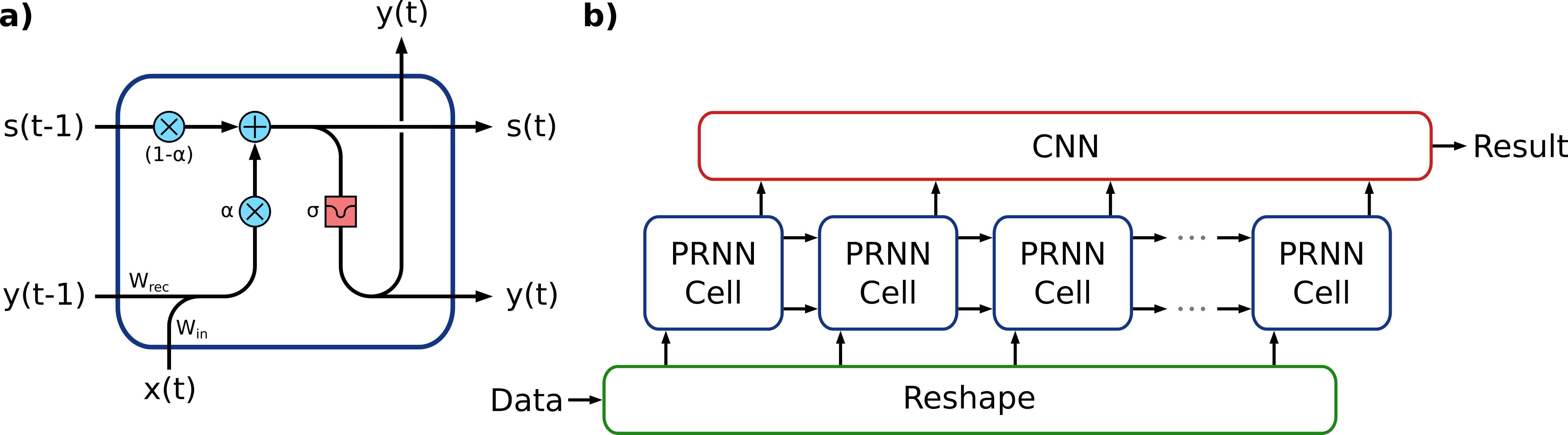}
    \caption{Framework of the discrete version of the photonic RNN model. (a) Schematic diagram of the discrete photonic RNN cell. Here $\alpha=\Delta t/ \tau$ (b) An example of connecting a photonic RNN to a CNN classifier for a time series classification task.}
\label{fig:RNN_cell}
\end{figure}

More generally, we can add an external input signal $\vec{x}(t)$ to the photonic RNN, and the dynamical equation become:
\begin{align}\label{eq:PRNN_ODE}
   & \tau\frac{d\vec{s}}{dt} = -\vec{s}+f(\vec{s}, \vec{x}, \vec{\theta}) \nonumber \\
   & f(\vec{s}, \vec{x}, \vec{\theta}) = {\bf{W}_{in}}\vec{x}(t) + {\bf{W}_{rec}}\sigma(\vec{s}(t)) + \vec{b},
\end{align}
where $\theta$ represents the trainable parameters for the network such as the bias $\vec{b}$, the recurrent photonic weight ${\bf{W}_{rec}}$, and the input weight ${\bf{W}_{in}}$. In this work, we approximate the dynamical equation in Eq.~\ref{eq:PRNN_ODE} using the {\it{forward-Euler method}} and use this discrete version of dynamics to construct the photonic RNN (PRNN) cell shown in Fig.~\ref{fig:RNN_cell} (a) and implement it on an FPGA.
The PRNN dynamics can be expressed by the following formula:
\begin{align}
    \tau\frac{d\vec{s}}{dt} & = -\vec{s}+f(\vec{s}, \vec{x}, \vec{\theta}) \Rightarrow \nonumber \\
{\vec{s}}(t+\Delta t) & = {\vec{s}}(t) +\Delta {\vec{s}} = {\vec{s}}(t) + \frac{\Delta t}{\tau}(-{\vec{s}}(t)+f(\vec{s}, \vec{x}, \vec{\theta})), \nonumber \\
& = (1-\frac{\Delta t}{\tau})\vec{s}(t) + \frac{\Delta t}{\tau}f(\vec{s}(t), \vec{x}(t), \vec{\theta}) \nonumber \\
& = (1-\alpha)\vec{s}(t) + \alpha f(\vec{s}(t), \vec{x}(t), \vec{\theta})
\label{eq:ode_approx}
\end{align}
where $\alpha=\Delta t/\tau$. The state variable ${\vec{s}}$ is updated based on Equation~\ref{eq:ode_approx} and can be constructed using Pytorch nn.module~\cite{Paszke2019PyTorchAI}. For the rest of the paper, we set $\Delta t/\tau=0.5$, and use an experimentally measured transfer function of a photonic neuron, $\sigma(x)=x^2/(x^2+(0.3+0.25x)^2)$.

\subsection{Neural Network Architecture and Experimental Settings}\label{sec:setting}
\begin{figure}[ht!]
    \centering
    \includegraphics[width=\linewidth]{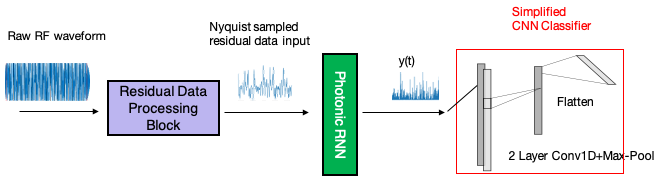}
    \caption{Procedure of the proposed RF fingerprinting system}
\label{fig:procedure}
\end{figure}

We propose an RF fingerprinting system which consists of a PRNN layer with PRNN unit cells and a CNN with two Conv1D layers and one fully connected layer. It is shown in Fig.~\ref{fig:RNN_cell} (b).
The input data has 64 channels with length 32 as described in Section~\ref{sec:data_form}, and the PRNN has 16 neurons to generate 16 channels of output sequence. The output $\vec{y}(t)$ from the PRNN is then sent to the convolutional layers.
We flatten the output of the convolutional layers and connect it to the fully-connected layer with 30 neurons and a log softmax activation function. The details of the parameters of the model are given in Appendix C, and the overall procedure to demonstrate RF fingerprinting using the PRNN-CNN model is shown in Fig.~\ref{fig:procedure}.

This architecture has 6,302 parameters in total, which is 50 times fewer than the number of parameters used in the model proposed in Ref.~\cite{Merchant2018DeepLF}. With the update equation provided in Equation~\ref{eq:ode_approx}, this PRNN can be trained end-to-end with the CNN model and with a back propagation through time (BPTT) algorithm~\cite{Werbos1990BackpropagationTT}. In this work, the training objective is the negative log likelihood loss given by the output of the model and targeted label. Glorot uniform initialization~\cite{Glorot2010UnderstandingTD} was used for initialization of all convolutional and dense layers, and Kaiming initialization~\cite{He2015DelvingDI} was used to initialize the parameters of PRNN layer. We use an ADAM optimizer~\cite{Kingma2015AdamAM} with an initial learning rate of 0.001. For each epoch, we train the model with a training dataset consisting of randomly shuffled data units and set the batch size to be 1,700. We validate the model by measuring its accuracy on the development dataset using $N=17$ segments of each transmission. When the classification accuracy on the development dataset doesn't increase in 10 consecutive epochs, we decrease learning rate by 50\% and keep training until 100 epochs.
\section{Experiments}\label{sec:exp}
In this work, we performed two sets of experiments.
Firstly, we implemented the baseline NRL model and our own using a NVIDIA GeForce GTX 1080 Ti GPU for training, validation, and testing. This experiment  was implemented in Python using Pytorch 1.4.0~\cite{Paszke2019PyTorchAI} as the back end and focused on comparing the classification performance of the two models. The results are shown in Sec.~\ref{sec:valid_exp}. Secondly, we performed real-time RF fingerprinting by implementing our model on a small-scale FPGA board, the PYNQ-Z1. Our method of realizing real-time classification on an FPGA and the results are summarized in Sec.~\ref{sec:fpga}. The hardware performance analysis is provided in Sec.~\ref{sec:perf}.
\subsection{Results}\label{sec:valid_exp}
\begin{figure}
	\includegraphics[width=\columnwidth]{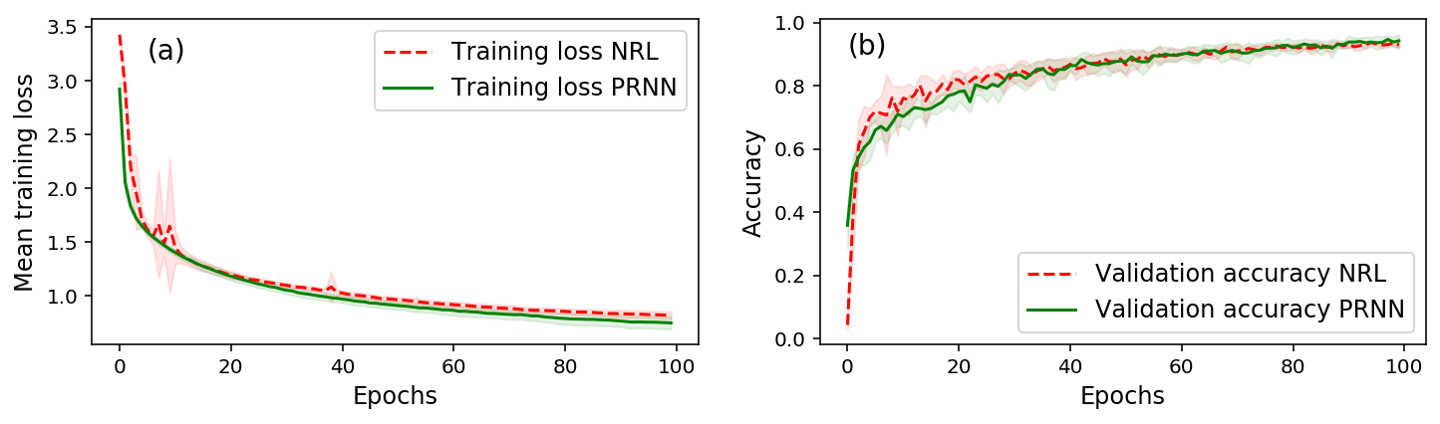}
	\caption{(a) Training loss. (b) Accuracy on the development dataset. The training and validation process is repeated 10 times, and the standard deviation is shown as the filled region. Red: Baseline NRL CNN; Green: Proposed PRNN-CNN model.}
	\label{fig:train_val}
\end{figure}

\begin{figure}[ht!]
    \centering
    \includegraphics[width=\linewidth]{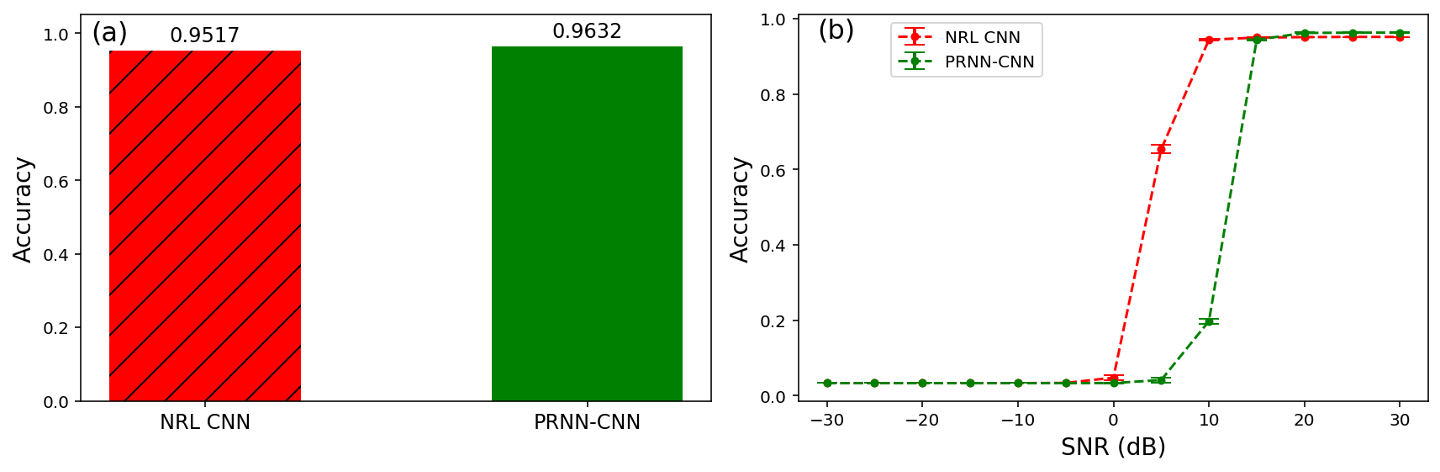}
    \caption{Comparison of NRL's classifier~\cite{Merchant2018DeepLF} and our PRNN-CNN classifier based on (a) maximum accuracy and (b) noise performance.}
\label{fig:result_comp}
\end{figure}

The training and evaluation results can be visualized in Fig.~\ref{fig:train_val}. We trained the PRNN-CNN model and compared it with the baseline NRL CNN model. To ensure the consistency, both models were trained 10 times. The standard deviation of the training loss and validation accuracy are shown as the filled region in Fig.~\ref{fig:train_val}. When using the baseline NRL CNN architecture, we achieve $95.17\%$ accuracy. The network has three convolutional layers and three fully connected layers with a total of 322,602 trainable parameters as detailed in Appendix C. We verified that our model converges to similar mean training loss and validation accuracy as the NRL CNN even with 50 times fewer parameters. 
For both the baseline and our proposed model, we  selected the trained parameters with the best accuracy on the development dataset and evaluated the models' performance with these parameters by measuring their classification accuracy on the test dataset. Both models achieve over 95\% classification accuracy on the test dataset as shown in Fig.~\ref{fig:result_comp} (a).

To further analyze the performance of the trained fingerprint classifier under less than ideal conditions, we added noise to both the I and Q channels of the test dataset to create variations in the input data. For simplicity, a standard artificial white Gaussian noise (AWGN) channel was implemented and simulated. In each run of the experiment, we added AWGN with a specific SNR in the set of $\{-30,-25,-20,...,25,30\}$(dB), and repeated for 20 times to check consistency of classification results. As shown in Fig.~\ref{fig:result_comp} (b), for the PRNN-CNN model, the classification accuracy stays above 95\% when the SNR is at least 15 dB while degrading when the SNR is below 15 dB. On the other hand, the NRL CNN model is more robust to the noise, as the accuracy starts decreasing when the SNR is below 10 dB. To improve the robustness to noise, Ref.~\cite{Lechner2018NeuronalCP} suggests using a liquid time constant RNN model to adjust the decay time of each neuron based on input perturbation. In this work, we mainly focus on constructing a compact network that can be implemented on an FPGA to provide low-latency and high-bandwidth processing, and we will leave improvement in noise performance to our future work.

\subsection{FPGA implementation}\label{sec:fpga}
In the second set of experiments, we sent the PRNN-CNN model with the trained parameters selected from Sec.~\ref{sec:valid_exp} to an FPGA to demonstrate RF fingerprinting in real-time.
The PRNN-CNN classifier was implemented on the low-cost PYNQ-Z1 FPGA board, chosen to demonstrate the hardware and cost minimization opportunities offered by the compact model. The FPGA design was created using the High-Level Synthesis (HLS) feature of the Vivado Design Suit~\cite{Vivado_HLS}, which allows a simplified C++ model of the design to be compiled first into a hardware description language format and then into a bitstream file for the FPGA.

\begin{figure}
	\includegraphics[width=\columnwidth]{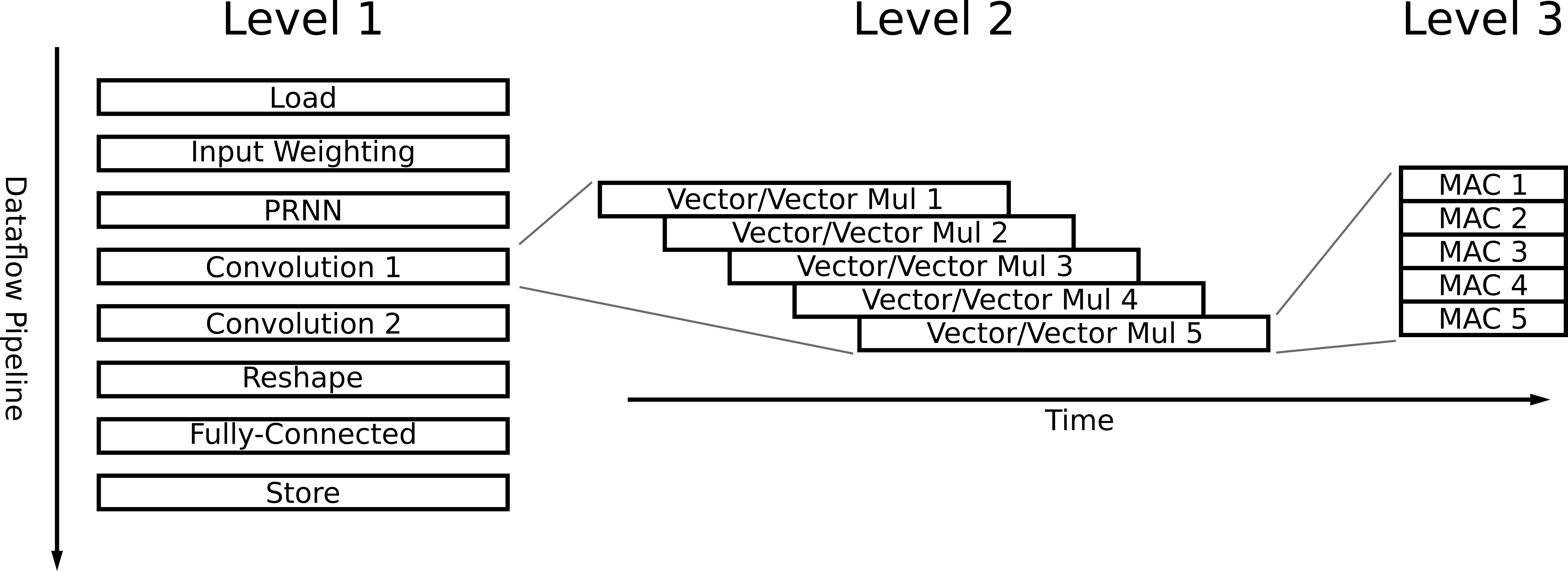}
	\caption{Design hierarchy for the FPGA implementation of the PRNN-CNN classifier.}
	\label{fig:hier}
\end{figure}

The FPGA design is organized in the three-level hierarchy shown in Figure \ref{fig:hier}. The first level consists of a series of stages within an HLS Dataflow pipeline. Each stage corresponds to a layer of the classifier, and a series of independent classifications pass through this pipeline, with an initiation interval corresponding to the latency of the slowest stage.

At the next level of abstraction, the operations within each stage are organized as a conventional HLS Pipeline. The execution path of each pipeline typically corresponds to a single vector-vector multiplication. A layer may be divided into a series of operationally identical such vector multiplications that progress through the pipelined execution path with a typical initiation interval of one cycle. (Read-after-write dependencies within the PRNN pipeline restrict the initiation interval to 16 cycles.)

At the final level, the execution path of each layer may be divided at points into parallel paths executed simultaneously. These parallel paths, consisting of individual multiply-accumulate (MAC) operations, allow for better utilization of the available digital signal processing resources and better minimization of overall latency.

\begin{figure}[ht!]
    \centering
    \includegraphics[width=\linewidth]{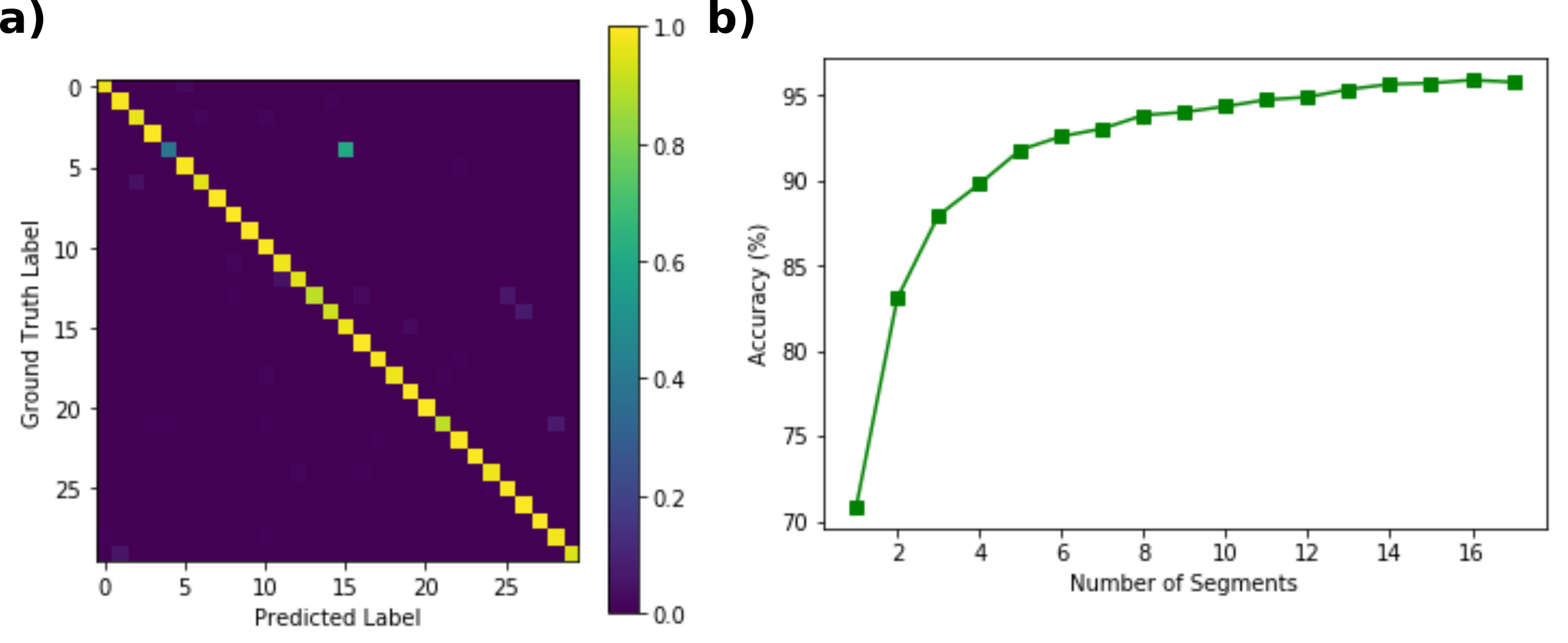}
    \caption{Performance of classification on 30 devices. (a) Confusion matrix of classification results. In the confusion matrix,  component $M_{ij}$ represents the ratio between the output prediction and ground truth label, i.e. $M_{ij}=N_i^{pre}/N_j^{tar}$, where $N_i^{pre}$ is the number of predictions for the $i$-th device, and $N_j^{tar}$ is the number of transmissions for the $j$-th device. The color yellow represents the perfect ratio, and purple represents the zero ratio.  (b) Impact of using different numbers of segments for classification. }
\label{fig:confusion}
\end{figure}

After constructing the PRNN-CNN with trained parameters on the PYNQ-Z1, we implemented RF fingerprinting on the test dataset. The results when using the PRNN-CNN model are shown as a confusion matrix in Fig. \ref{fig:confusion}(a). Our model reaches 95.90\% accuracy. There was only a 0.42\% decrease for the FPGA implementation due to slight additional sources of imprecision.

During testing there is a trade-off between accuracy and throughput. Maximum accuracy requires that all 17 segments of a transmission be used, but using only 8 segments still allows for a high level of accuracy, nearly 94\%, while more than doubling the throughput due to the reduction in calculation per classification required. The precise relationship between number of segments and accuracy is shown in Figure \ref{fig:confusion}(b).

\subsection{Performance Analysis}\label{sec:perf}

\begin{table}[h]
\centering
\renewcommand\arraystretch{1.3}
\fontsize{10pt}{10pt}\selectfont
{
\begin{tabular}{c c c c}
\hline
\hline
\textbf{Metric} & \textbf{NRL Model} & \textbf{PRNN-CNN Model} & \textbf{Improvement Factor}\\
\hline
\hline
Energy (\si{\micro\joule}) & 6,194 & 15 & 413 \\
Latency (\si{\micro\second}) & 26,190 & 219 & 119\\
Throughput (1/s) & 50 & 12,192 & 244\\
\hline
\end{tabular}}
\caption{Estimated NRL and PRNN-CNN classifier performance results on per-classification basis. Timing results for the PRNN-CNN confirmed experimentally.}
\label{tab:performance}
\end{table}

Table \ref{tab:performance} shows the performance estimations for FPGA implementations of the NRL and PRNN-CNN classifiers. Both classifiers are built with the same hierarchical design principles, but the NRL network could not be implemented on the PYNQ-Z1 board due to insufficient available memory to hold the parameters and intermediate data representations. In order to estimate energy per classification, consumption was assumed to scale linearly with the number of multiply-accumulate (MAC) operations per inference, with 170 pJ per MAC assumed for an energy-efficient FPGA \cite{Giefers2016MeasuringAM}. Latency and throughput estimates are derived directly from Vivado HLS compilation reports.

\begin{figure}
    \centering
    \includegraphics[width=0.75\linewidth]{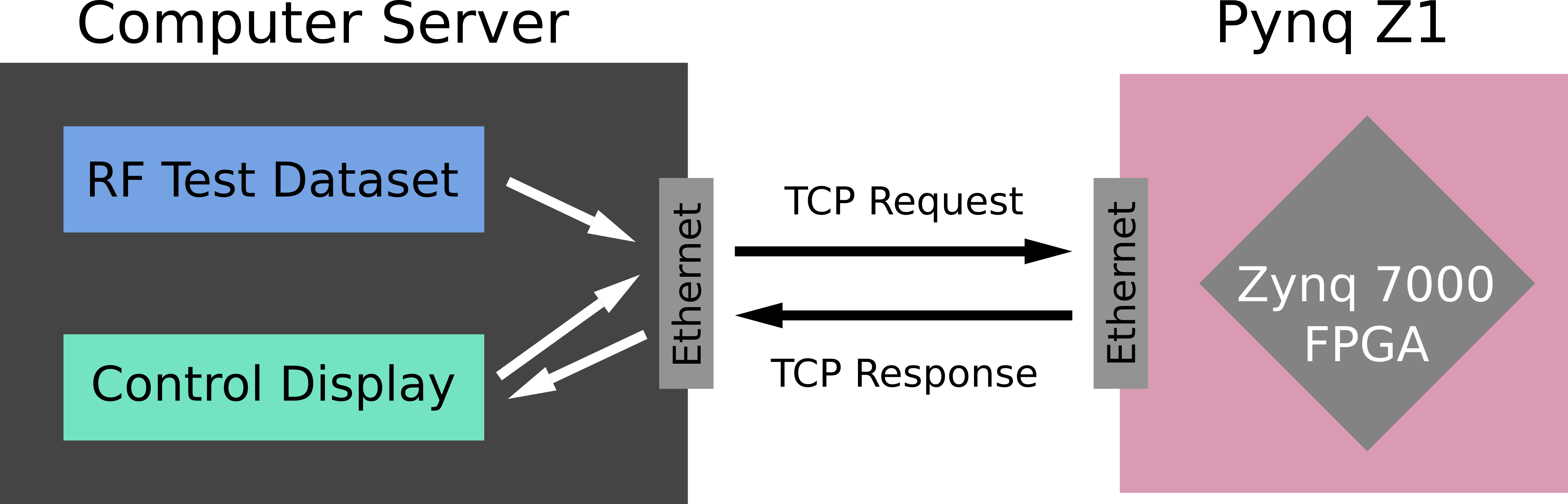}
    \caption{Diagram of the real-time test setup.}
\label{fig:demoSetup}
\end{figure}

In order to confirm the timing behavior of the system, the PRNN-CNN model was implemented on the PYNQ-Z1 and evaluated experimentally in a real-time test. As shown in Figure \ref{fig:demoSetup}, randomly selected radio frequency data segments were streamed from a remote computer to the FPGA board, which processed them in real time. The measured timing showed negligible deviation from the estimated values.

The reduction in size and simplification of the PRNN-CNN network with respect to the NRL network allows for a 413-times reduction in power consumption, a 119-times reduction in latency, and a 244-times increase in throughput based on the estimated parameters. Latency reduction and throughput maximization are key to enabling real-time processing, as both must fall within context-dependent thresholds to prevent data queuing and to enable fast response. Improvement in these metrics allows real-time RF fingerprinting to be expanded to environments necessitating very fast response to large-bandwidth RF data.

\section{Conclusions}\label{sec:conclusion}
In this work, we have demonstrated the following criteria for a real-time RF fingerprinting system:
\begin{itemize}
  \item Accuracy improvement: We propose a novel PRNN-CNN model for RF fingerprinting, which achieves over a 96\% accuracy when the SNR is at least 15 dB.
  \item Compact model: The proposed model has been shown to reduce the number of parameters by 50 times compared to the large CNN model proposed in Ref.~\cite{Merchant2018DeepLF}.
  \item Implementability: The PRNN-CNN model can be fully implemented on a small-scale FPGA, and has the potential to be implemented on a hybrid photonic/FPGA edge processor that consists of a photonic RNN to process sequential data with low latency and a CNN on the FPGA to enable massive parallel machine learning inference.
  \item Hardware efficiency: We estimate the power and latency for the proposed RF fingerprinting system, and show the power consumption may be improved by over 400
  times and the latency reduced to more than 100 times 
  over previous approaches.
\end{itemize}
With the above criteria, we show the potential of using the small-scale FPGA processor to perform RF fingerprinting in real time. A number of issues are still left open in this investigation such as the improvement for noisy data, and real-time implementation of RF fingerprinting on a hybrid photonic/FPGA processor. We will leave these topics as our future work.

\section*{Broader Impact}

High-accuracy, real-time signal classification on a hardware platform with small size, weight and power (SWaP) will enable the deployment of ML technique at the wireless communication edge. Our paper makes important contributions to this endeavor with both an innovative compact neural network architecture and experimental testing results of RF fingerprinting. Within the wireless communication domain, our work can contribute to fast modulation format classification (skipping the residual data generation used in RF fingerprinting), and RF emitter positioning. Fundamentally, time series data classification is one of the challenging problems facing data science, and our proposed photonic-circuits-inspired compact network achieves fast processing with a very small scale FPGA and has the potential to impact much broader areas. However, if our proposed model can be applied to decode contents from RF signals, then it could lead to a potential negative societal impact.

\section*{Acknowledgment and disclosure of funding}
The authors would like to thank Thomas Carroll and Bryan Nousain from Naval Research Lab for providing the data source and discussions on RF fingerprinting techniques, and Dr. Young-Kai Chen and Dr. Gregory Jones from DARPA for the support and insightful discussions. This work is financially supported by DARPA PEACH program (AWD1006165).

\bibliographystyle{unsrtnat}
\bibliography{neurips_2021}

\newpage
\appendix

\section*{Appendix}
In the supplemental material, we will provide details on residual data processing, neural networks on silicon photonic circuits, and the NRL and PRNN models for RF fingerprinting. Our live demonstration for real-time RF fingerprinting using PYNQ-Z1 FPGA board is available at \url{https://www.youtube.com/watch?v=CIfddiscE3k}.

\section{Residual data processing}

The RF transmitters of the same type follow the same steps and communication protocols to process the transmitting bits by encoding, modulation, pulse shaping and up-conversion to generate the signals for transmission. Due to the electronic component's performance variations, the transmitted signals carry small and hidden information which is intrinsic to each individual transmitter. We employ a residual data processing step, which extracts an error signal by subtracting the recovered data waveform from the received raw data waveform. The error signal, instead of the received raw data, is fed to the learning algorithm and neural networks for classification. Ideally, the error signal contains only residual signal elements directly associated with the intrinsic makeup of the transmitting device, rather than the transmitted bit information which the transmitters can change almost randomly.

 The residual data processing, which serves as a pre-processing step before the neural network classification, removes the information data from the signal by recovering the measured data, generating an ideal signal matching this data, then subtracting the ideal signal from the measured signal, leaving only nonidealities associated intrinsically with each individual device. In order to generate the ideal signal, the measured signal is decoded then re-encoded. This section describes how this was done.

\subsection{Raw RF data settings}
The Naval Research Laboratory collected radio frequency data measured from thirty ZigBee Pro devices operating at a central carrier frequency of 2.405 GHz, and a power level of 18dBm. The data provided has been downconverted to a complex baseband signal at 16 MHz in a 32-bit floating point form (4 bytes per value, or 8 bytes per complex sample). No further corrections had been applied to the data. In the following subsections, we will describe the details of residual data processing step by step.

\subsection{Separating Out Transmissions}

\begin{figure}
  \includegraphics[width=0.75\textwidth]{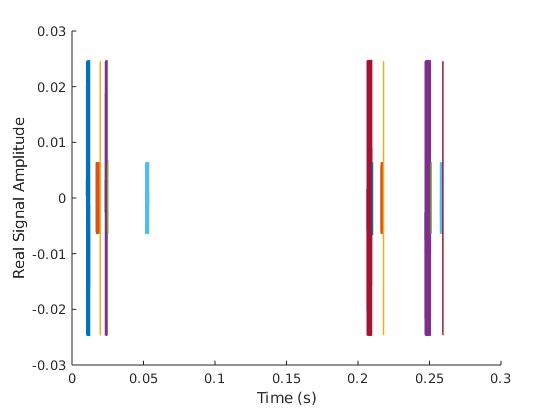}
\centering
\caption{The real portion of the first 300 ms of broadband radio frequency data for a single device. Each color corresponds to a separate transmission.}
\label{fig:multipleTransmissions}
\end{figure}

Under the ZigBee protocol, a device broadcasts data separated into individual transmissions, each of which may be considered separately. These are sliced out by identifying large duration during which the signal is below a threshold, which separate transmissions. A segment of data with multiple transmissions is shown in Figure \ref{fig:multipleTransmissions}.

\subsection{Trimming and Synchronizing the Transmission}

To be decoded correctly, transmissions must be aligned precisely, trimming off the start and end of the transmission to begin and end at the right sample. Each transmission begins with a fixed five bytes preamble, corresponding to four bytes of zeros and one byte equal to 0xA7. The beginning of the transmission may be identified by matching the signal corresponding to an ideal preamble to the measured signal. This is complicated by the fact that the initial phase of the signal is unknown. To avoid this issue, the discrete derivative of the phase of the complex ideal and measured signals were compared (unwrapped to avoid issues of phase jumping between $-2\pi$ and $2\pi$). This quantity may still be accurately matched to the preamble while being independent of the initial phase.

Mathematically $t_0$ was determined such that the following quantity was minimized:
\[
  \sqrt{\sum_t \left\{   \frac{d}{dt} \left[ \mathrm{unwrap}(\angle x(t+t_0)) \right]- \frac{d}{dt} \left[ \mathrm{unwrap}(\angle y(t)) \right] \right\}}
\]
Where $\angle x$ represents the phase of the measured signal and $\angle y$ the phase of the reference preamble, with the sum over the length of $y$.

\begin{figure}
  \includegraphics[width=0.75\textwidth]{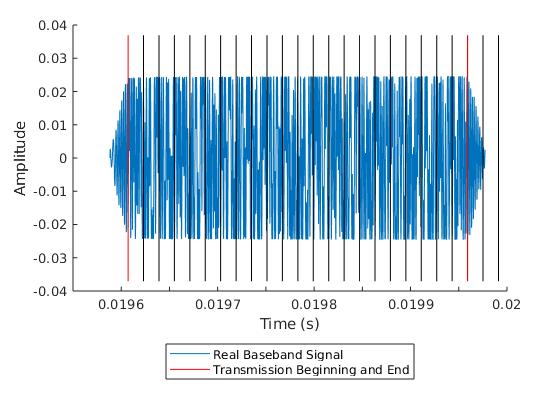}
\centering
\caption{The real portion of a single transmission, with beginning and end determined and the entire transmission segmented into separate samples.}
\label{fig:untrimmedTransmission}
\end{figure}

Figure \ref{fig:untrimmedTransmission} shows a single transmission, equivalent to the third transmission (in yellow) in Figure \ref{fig:multipleTransmissions}. The plot is segmented into signals (four bytes sequences) separated by the vertical black and red lines. The first ten signals correspond to the preamble. (This is a very short transmission, and the preamble makes up the bulk of it). The position of the first red line, the start of the transmission, was determined through the above process.

Another red line indicates the end of the transmission. Every transmission has a length equivalent to an even number of signals, and the end is determined by finding the latest sample obeying this requirement. As the signal decoder is reasonably robust to additional length in the case that the chosen end sample is too late, more precision is not required.

\subsection{Correcting Phase and Frequency Offsets}

Data transmitted according to the ZigBee protocol is encoded using offset quadrature phase shift keying (O-QPSK). The signal has in-phase (I) and quadrature (Q) components which must be accurately separated.

The following model on the separation of the I and Q components is drawn from a paper published by ~\citet{Merchant2018DeepLF}, with some adjustments and additions.

The data is encoded by the transmitter as a complex baseband signal $\tilde{a}(t)$, with time $t$. The transmitted signal $b(t)$, with carrier frequency $\omega_c$ would then be:
\begin{equation}
  b(t)=\mathrm{Re}\left[ \tilde{a}(t)e^{i\omega_c t} \right]
\end{equation}
However, suppose the transmitter frequency and phase differs from the receiver frequency and phase. Modeling the difference as a phase offset $\phi_o$ and frequency offset $\omega_o$ at the transmitter side, the transmitted signal is then:
\begin{equation}
  b(t)=\mathrm{Re}\left[ \tilde{a}(t)e^{i\left[ \left( \omega_c+\omega_o \right)  t +\phi_0\right] } \right]
\end{equation}
The receiver measures the signal and downconverts it to a baseband signal $\tilde{c}(t)$, "removing" the carrier frequency $\omega_c$. In addition, there is a substantial attenuation in signal amplitude between the transmitter and receiver, represented by $\alpha$, and other phase distortion factors, small compared to the effect of $\omega_o$ and $\phi_o$, represented by $D(t)$. $D(t)$ includes elements associated with a device's unique wireless signature.
\begin{equation}
  \tilde{c}(t)=\alpha\tilde{a}(t)e^{i\left( \omega_o t+\phi_0+D(t) \right) }
\end{equation}
The challenge is to extract $\tilde{a}(t)$ (or, more accurately, its phase for QPSK) from the measured $\tilde{c}(t)$ so that its effect may be subtracted out to produce the error signal, incorporating effects associated with $D(t)$. Let $\omega$ and $\phi$ represent the estimated phase and frequency correction, designed to compensate for $\omega_o$ and $\phi_o$. These are applied to the measured signal $\tilde{c}(t)$ to produce a corrected signal $\tilde{d}(t)$:
\begin{equation}
  \tilde{d}(t)=\tilde{c}(t)e^{-i(\omega t+\phi)}
\end{equation}
We will find $\omega$ and $\phi$ such that $\omega=\omega_o$ and $\phi=\phi_0$:
\begin{equation}
  \tilde{d}(t)=\tilde{c}(t)e^{-i(\omega t+\phi)}=\alpha\tilde{a}(t)e^{i\left[ (\omega_o-\omega)t +\phi_o-\phi +D(t)\right] }=\alpha\tilde{a}(t)e^{iD(t)}
\end{equation}
Representing $a(t)$ and $c(t)$ as quantities with amplitudes and phases, the latter represented $\phi_a(t)$ and $\phi_c(t)$, we have a phase equation:
\begin{equation}
  \phi_c(t)-\omega t -\phi=\phi_a(t)+D(t)\to \mathrm{unwrap}\left( \phi_c(t)-\phi_a(t) \right) =\omega t+\phi+D(t)
  \label{eq:regression}
\end{equation}
The function unwrap() above means that the phase difference is unrolled, allowing it to increase past $\pi$ to $2\pi$, $3\pi$, $4\pi$ and so forth rather than circling back to $-\pi$.

Equation \ref{eq:regression}, in words: the phase difference between the measured signal $\tilde{c}(t)$ and the ideal signal $\tilde{a}(t)$ increases roughly linearly, with an initial phase offset $\phi$ and a slope corresponding to the frequency offset $\omega$. There is also a small deviation from this linear model, $D(t)$. If one has both the measured signal and an idea signal, linear regression may be used to estimate $\omega$ and $\phi$.

The fixed first five bytes of a transmission (the preamble) are used to generate an ideal reference signal matching the first five bytes of the measured transmission. Linear regression produces estimates of $\omega$ and $\phi$. In theory, these corrections may be applied to the measured signal $\tilde{c}(t)$ to produce $\tilde{d}(t)$, which may be decoded, as its I and Q components are appropriately separated.

Unfortunately, the estimates of the frequency offset $\omega$ based on the first five bytes of the signal are too imprecise. As time progresses, any slight inaccuracy in $\omega$ caused by the semi-random $D(t)$ builds up. Once the phase discrepancy at a certain point in time gets too high, the I and Q components of the signal will not be accurately separated and the decoder fails.

\begin{figure}
  \centering
  \begin{subfigure}{0.45\textwidth}
\centering
  \includegraphics[width=\textwidth]{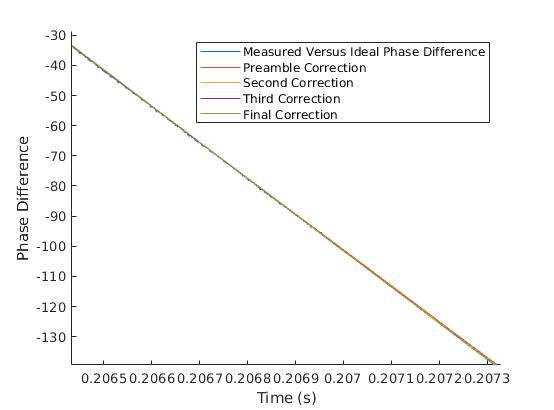}
\caption{Early in the transmission. Note that all corrections are accurate.}
\label{fig:unshiftedEarly}
\end{subfigure}\hfill
\begin{subfigure}{0.45\textwidth}
\centering
 \includegraphics[width=\textwidth]{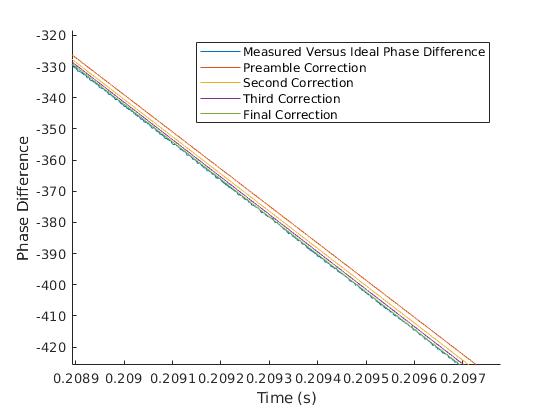}
\caption{Near the end of the transmission. The earlier corrections have drifted substantially.}
\label{fig:unshiftedLate}
\end{subfigure}
\caption{The phase difference between a typical measured transmission and each iterative linear correction.}
\label{fig:unshifted}
\end{figure}

Figure \ref{fig:unshifted} illustrates this problem. Early in the transmission, shown in Figure \ref{fig:unshiftedEarly}, the linear fit associated with the preamble correction is quite accurate. By the end of the transmission, in Figure \ref{fig:unshiftedLate}, the fit has drifted several radians from the phase difference.

The solution was to implement an iterative correction algorithm. The preamble correction is used to decode the signal up to the point that the decoder fails. The correctly decoded portion may then be used to generate a longer ideal signal and therefore a longer phase difference plot. Linear regression on this extended dataset produces more precise estimates of the phase and frequency offset, which in turn allow the measured signal to be more accurately corrected. This loop continues until the entire signal had been decoded, typically after 1-3 iterations following the preamble correction.

\begin{figure}
  \centering
  \includegraphics[width=\textwidth]{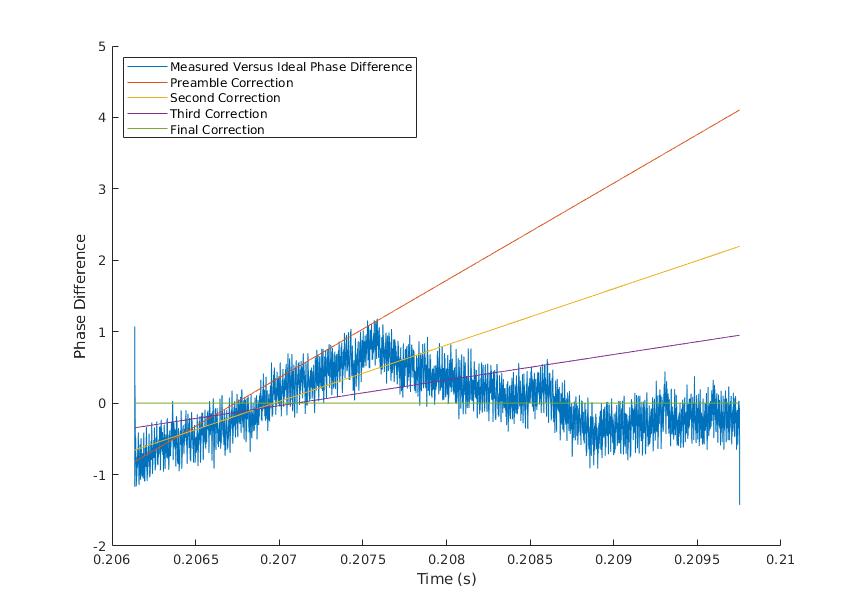}
  \caption{The phase difference between a typical measured transmission and each iterative correction, with the final correction subtracted out.}
\label{fig:shifted}
\end{figure}

Figure \ref{fig:unshifted} shows each of these further corrections. Notably, in Figure \ref{fig:unshiftedLate} the final correction has not drifted from the phase difference signal. Figure \ref{fig:shifted} illustrates this in a more readable fashion. The final correction has been subtracted from the phase difference and the other corrections, so that the final correction itself appears flat. Each correction represents the best fit for a segment of the data beginning at the start of the transmission, with this segment lengthening with each correction.

With the entire signal decoded, we have a corrected signal $\tilde{d}(t)$ and an idea signal corresponding to $\alpha \tilde{a}(t)$, the original ideal signal scaled to match the measured signal. The error signal $\tilde{e}(t)$ is then:
\begin{equation}
  \tilde{e}(t)=\tilde{d}(t)-\alpha \tilde{a}(t)
\end{equation}

\begin{figure}[ht!]
    \centering
    \includegraphics[width=1.0\linewidth]{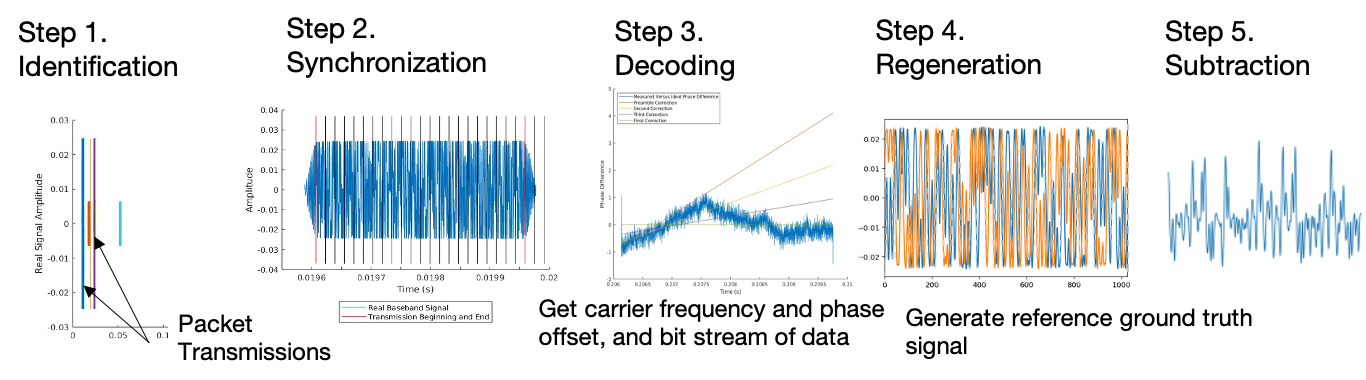}
    \caption{Procedure of residual data processing}
\label{fig:residual_proc}
\end{figure}
The overall residual processing is summarized in Fig.~\ref{fig:residual_proc}.
\section{Neural networks on silicon photonic integrated circuits}
\subsection{Mapping neuronal function on silicon photonic integrated circuits}
\begin{figure}[th]
    \centering
    \includegraphics[width=1.0\linewidth]{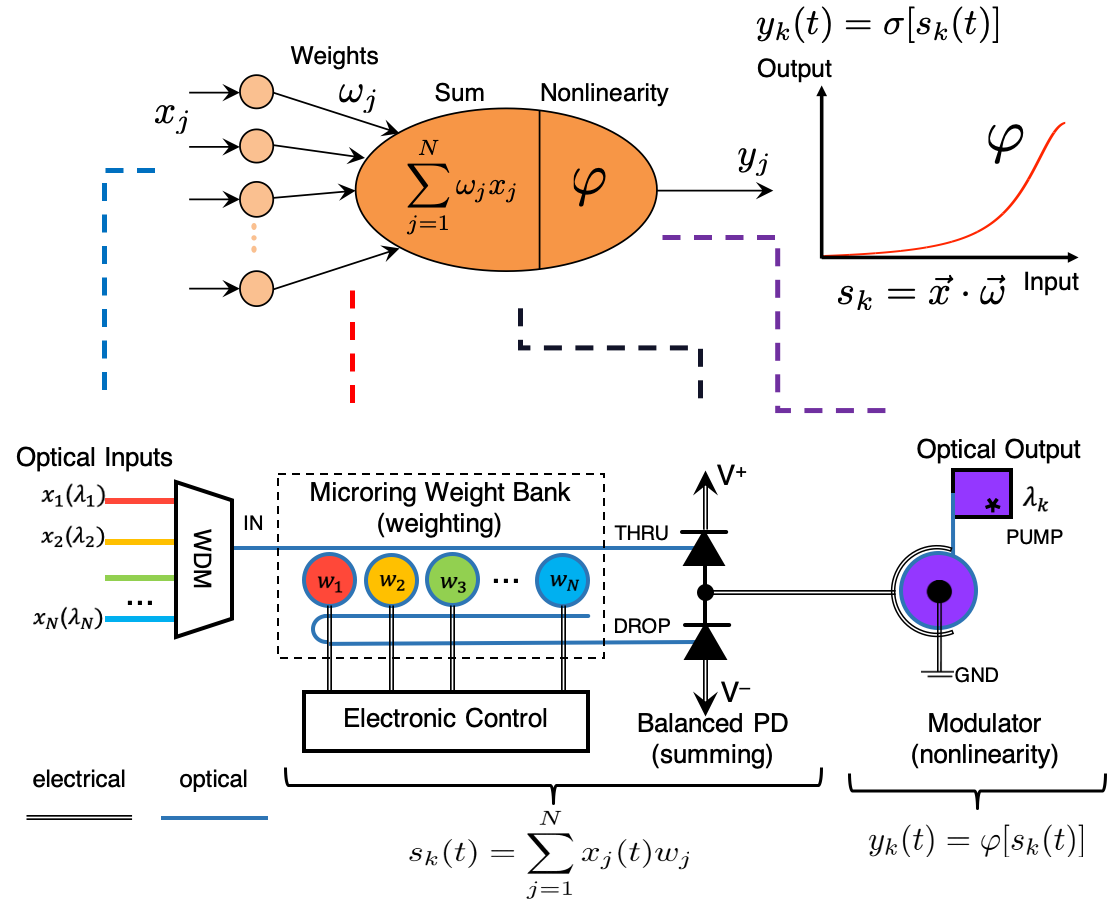}
    \caption{Mathematical abstraction of neuronal function and silicon photonics implementation.}
\label{fig:schematics}
\end{figure}

The PRNN cell introduced in Section~\ref{sec:PRNN-CNN} is inspired by silicon photonic neural networks. In this section, we will introduce the details of this model, and show how it is implemented on a photonic integrated circuit. The neural information processing of each neuron can be decomposed into four parts: (1) input signals receiving, (2) independent weighting of inputs, (3) summation, and (4) nonlinear transformation. These operations can be mapped and implemented on a silicon photonic integrated circuit as shown in Fig.~\ref{fig:schematics}. The input signals are encoded in the power of the optical sources. Each optical input has a unique wavelength to represent the signal from a different pre-synaptic neuron. These signals are multiplexed through wavelength-division multiplexing (WDM) to an optical fiber, which is then coupled to an on-chip waveguide. The weighting is implemented using a set of silicon micro-ring resonators (MRR) ~\cite{Tait:16anal,Tait:16multi}. Each of the micro-ring resonators is designed to resonate with each input wavelength and control its weight independently by fine tuning the resonance to change the transmission. The weighted optical signals are summed by balanced photodetectors (BPDs), which linearly transform the sum of the optical signals to \emph{photocurrent}. The photocurrent is then sent to a silicon micro-ring modulator (MRM) ~\cite{Tait:2019} as input to a neuron. It is worth noting that the weights are in the range of [-1,1], so usually an electrical amplifier such as a transimpedance amplifier (TIA) will be added to scale up the weighted value. Due to the carrier-induced nonlinear effect, a MRM will provide the nonlinear activation to the optical pump signal. This mechanism shows the working principal of a neuron node on a silicon photonic circuit. The optical output of the MRM can be further sent to other photonic neuron nodes to form a network system.
\subsection{Implementation of a PRNN node}
\begin{figure}[ht!]
    \centering
    \includegraphics[width=1.0\linewidth]{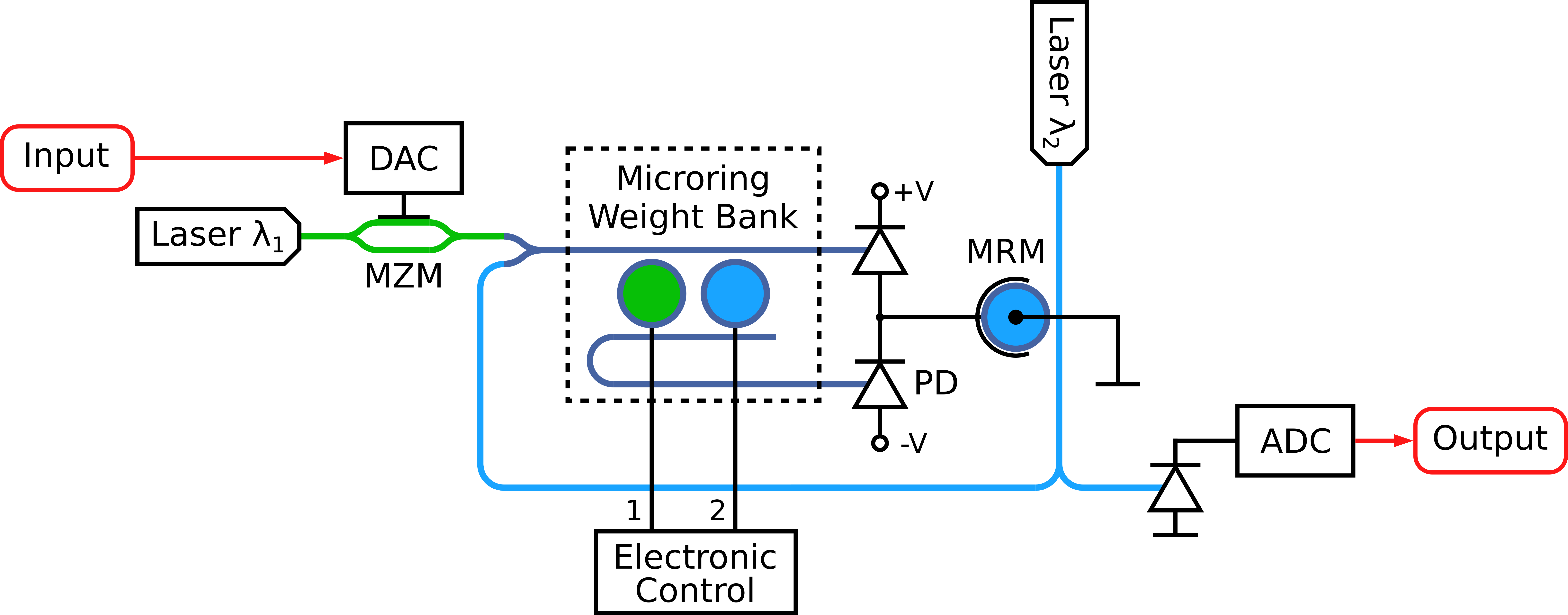}
    \caption{Photonic RNN node implemented with microring resonators. DAC: digital-to-analog converter, ADC: analog-to-digital converter, MZM: Mach-Zehnder modulator, PD: photodetector, MRM: microring modulator.}
\label{fig:PRNNnode}
\end{figure}

In Fig.~\ref{fig:PRNNnode}, we show the basic architecture of a recurrent circuit of a silicon photonic neuron with MRR and MRM. To implement this PRNN node, one would convert the input digital data to an analog signal that drives a Mach-Zehnder modulator to generate an input signal in the optical domain with wavelength $\lambda_1$. Each of the MRRs has a different transmission coefficient for optical signals at different wavelengths. After passing the MRR weight bank, the optical signal is converted to an electrical signal that drives a micro-ring modulator to modulate the continuous-wave laser at wavelength $\lambda_2$. The $\lambda_2$ optical signal carrying the weighted information of the input signal is fed back to the MRR bank to form an RNN node. The dynamics of this system is the same as Eq.~\ref{eq:PRNN_ODE}, and the RNN terminology and photonics equivalent are given in Table.~\ref{tab:RNN_photonics}.

\begin{table}[ht]
\centering
\renewcommand\arraystretch{1.0}
\fontsize{10pt}{10pt}\selectfont
{
\begin{tabular}{c c c}
\hline
\hline
& \textbf{RNN terminology} & \textbf{Photonic equivalent}\\
\hline
\hline
$\tau$ & Time constant of post-synaptic node &  RC time constant of the link between BPDs and MRM \\
y & Output of post-synaptic node & Normalized output optical power of MRM \\
$\bf{W}_{in}$ & Input weight matrix & MRR weights associated with input wavelengths \\
$\bf{W}_{rec}$ & Recurrent weight matrix & MRR weights associated with output wavelengths \\
$\sigma$(x) & Nonlinear activation function & Nonlinear transfer function of MRMs  \\
$\vec{b}$ & Bias of pre-synaptic node & DC current injected to MRM \\
\hline
\\
\end{tabular}}
\caption{RNN terminology and photonics equivalent.}
\label{tab:RNN_photonics}
\end{table}

\section{Neural network classifiers for RF fingerprinting}\label{sec:NRL_CNN_RNN}
In Ref.~\cite{Merchant2018DeepLF}, the authors proposed a multi-layer convolutional neural network to classify the transmission of 7 identical ZigBee devices. We have built the same structure CNN as proposed in ~\cite{Merchant2018DeepLF} but for classification of 30 devices. The network has 322,602 trainable parameters, and Table ~\ref{tab:CNN1} shows the structure of multi-layer CNN used for classification of 30 Zigbee devices. In this CNN, aside from the last layer which has a log softmax nonlinearity, all the layers use the exponential linear unit as their activation function, which has the following nonlinearity:
\begin{align*}
ELU(x)=
    \begin{cases}
    x, \text{   if  } x > 0 \\
    \alpha (exp(x)-1), \text{  if  } x \leq 0 \\
    \end{cases}
\end{align*}
In our experiment, we set the hyperparameter $\alpha=1$.\\
\begin{table}[ht]
\centering
\renewcommand\arraystretch{1.0}
\fontsize{10pt}{10pt}\selectfont
{
\begin{tabular}{c c c c}
\hline
\hline
\textbf{Layer} & \textbf{Dimension} & \textbf{Parameters} & \textbf{Activation} \\
 & (channel, length) & & \\
\hline
\hline
Input  &  2$\times$1024 & $-$ & $-$\\
Conv1D & 128$\times$19 & 4992 & ELU\\
Max Pooling & $-$ & $-$& $-$\\
Conv1D & 32$\times$15 & 61472 & ELU\\
Max Pooling & $-$ & $-$ & \\
Conv1D & 16$\times$11 & 5648 & ELU\\
Max Pooling & $-$ &  $-$& \\
Flatten & $-$ & $-$ & \\
Fully-Connected & 128 & 239744 & ELU\\
Dropout & $-$ & $-$ & \\
Fully-Connected & 64 & 8256 & ELU\\
Dropout & $-$ & $-$ & \\
Fully-Connected & 30 & 1950 & Log Softmax\\
\hline
\\
\end{tabular}}
\caption{Multi-layer CNN architecture for 30 devices classification.}
\label{tab:CNN1}
\end{table}
The details of our proposed PRNN-CNN architecture are shown in Table~\ref{tab:PRNN}. It has a total of 6,302 parameters. In the PRNN layer, the nonlinear activation function is a Lorentzian function, which is designed to match the nonlinear transfer function of a silicon photonic modulator~\cite{Tait:2019}.
\begin{table}[ht!]
\centering
\renewcommand\arraystretch{1.3}
\fontsize{10pt}{10pt}\selectfont
{
\begin{tabular}{c c c c}
\hline
\hline
\textbf{Layer} & \textbf{Dimension} & \textbf{Parameters} & \textbf{Activation} \\
 & (channel, length) & & \\
\hline
\hline
I/Q Input  &  2$\times$1024 & $-$ & $-$\\
Reshape  &  64$\times$32 & $-$ & $-$\\
PRNN & 16$\times$32 & 1312 & Lorenztian\\
Conv1D & 16$\times$5 & 1296 & ELU\\
Max Pooling & $-$ & $-$& $-$\\
Conv1D & 16$\times$3 & 784 & ELU\\
Max Pooling & $-$ & $-$ & \\
Flatten & $-$ & $-$ & \\
Fully-Connected & 30 & 2910 & Log Softmax\\
\hline
\hline
\end{tabular}}
\caption{PRNN assisted RF fingerprinting system.}
\label{tab:PRNN}
\end{table}


\end{document}